\documentclass[tradiabstract]{aa}
\usepackage[dvipdf]{graphicx}
\usepackage{txfonts}
\input{psfig}
\newcommand{\ltsima} {$\; \buildrel < \over \sim \;$}
\newcommand{\gtsima} {$\; \buildrel > \over \sim \;$}
\newcommand{\lta} {\lower.5ex\hbox{\ltsima}}
\newcommand{\gta} {\lower.5ex\hbox{\gtsima}}
\newcommand{\kms}{km s$^{-1}$}
\newcommand{\HI}{H{\,{\sc i}}}

\begin{document}

\title{Is cold gas fuelling the radio galaxy 
NGC~315?\thanks{Based on observations  with the Westerbork Synthesis Radio Telescope
(WSRT), the Very Large Array (VLA) and Very Long Baseline Interferometer
(VLBI).}}

\titlerunning{Is cold gas fuelling the radio galaxy 
NGC~315?}
\authorrunning{Morganti et al.}  

\author{R. Morganti\inst{1,2}, A.B. Peck\inst{3,4},
 T.A. Oosterloo\inst{1,2},  G. van Moorsel\inst{4}
A. Capetti\inst{5},\\
R. Fanti\inst{6,7}, P. Parma\inst{6} \and H.R. de Ruiter\inst{6,8}}

\institute{Netherlands Foundation for Research in Astronomy, Postbus 2,
7990 AA, Dwingeloo, The Netherlands
\and
Kapteyn Astronomical Institute, University of Groningen, P.O. Box 800,
9700 AV Groningen, The Netherlands
\and
Joint ALMA Office, Av El Golf 40, piso 18, Santiago, 7550108 Chile
\and
National Radio Astronomy Observatory, Socorro,
             NM 87801, USA
\and
Osservatorio Astronomico di Torino, Strada Osservatorio 25,
I-10025 Pino Torinese, Italy
\and
INAF, Istituto di Radioastronomia, Via Gobetti 101, I-40129, Bologna, Italy
\and
Dipartimento di Fisica dell'Universit{\`a} di Bologna, Via Irnerio 46,
I-40126 Bologna, Italy
\and
Osservatorio Astronomico di Bologna, Via Ranzani, 1, I-40127 Italy}

\offprints{morganti@astron.nl}
\date{Received ...; accepted ...}

\abstract{We present WSRT, VLA and VLBI  observations of the \HI\ absorption in
the radio galaxy NGC~315. The main result is that {\sl two} \HI\ absorbing
systems are detected against the central region. In addition to the known highly
redshifted, very narrow component, we detect relatively broad (FWZI
$\sim$$150$ \kms) absorption. This broad component is redshifted by $\sim$$80$ \kms\
compared to the systemic velocity, while the narrow absorption is redshifted $\sim 490$ \kms.  Both \HI\ absorption components are
spatially resolved at the pc-scale of the VLBI observations. The broad component
shows strong gradients in density (or excitation) and velocity along the jet. We
conclude that this gas is physically close to the AGN, although the nature of the gas resulting in the broad absorption is not completely clear. 
The possibility that it is entrained by the radio jet (and partly responsible of the deceleration of the jet) appears unlikely. 
Gas located in a thick circum-nuclear toroidal structure, with orientation 
similar to the dusty, circumnuclear disk observed with HST,  cannot be completely ruled out although it appears  difficult to reconcile with the observed morphology and kinematics of the \HI.  
A perhaps more likely scenario is that the gas producing the broad absorption could be (directly or indirectly) connected with the
fuelling of the AGN, i.e. gas  that is falling into the nucleus. If this is the case, the accretion rate derived is similar
(considering all uncertainties) to that found for other X-ray luminous
elliptical galaxies, although lower than that derived from the radio core
luminosity for NGC~315.   
The data also show that, in contrast to the broad component, the density
distribution of the narrow component is featureless. Moreover, in the WSRT
observations we do detect a small amount of \HI\ in emission a few kpc SW of the
AGN, coincident with faint optical absorption features and at velocities very
similar to the narrow absorption. This suggests that the gas causing the narrow
absorption is not close to the AGN and is more likely caused by  clouds  falling
into NGC 315. The environment of NGC 315 turns out to be indeed quite gas rich
since we detect five gas-rich companion galaxies  in the immediate vicinity of
NGC 315.}


\maketitle

\section{Introduction}
\label{intro}

\subsection{HI 21-cm absorption in low luminosity radio galaxies}

The channeling of the gas to the very inner regions of a galaxy is considered 
to be the mechanism that can transform a ``starving'' black hole into an active
nucleus. Mergers and interactions can play an important role in supplying the
fuel and providing the conditions for the gas to reach the centre (Wilson 1996).
However, the relationship with mergers is not one-to-one and recent studies of radio
galaxies have shown that the activity in some of these galaxies may be
associated instead with the {\sl slow}  accretion of (hot) gas (Best et al.\
2005, 2007; Croton et al.\ 2006).  In the case of radio-loud AGN, the way the
accretion of gas proceeds can have important implications in determining
the characteristics of the radio source that is associated with the AGN. In
particular, the difference between powerful, edge-brightened and low luminosity,
edge-darkened radio galaxies could reflect a change in the mode of accretion:
advective low efficiency/rate flow in the latter (Allen et al. 2006, Balmaverde,
Baldi \& Capetti 2008) and standard optically-thick accretion disks in the
former.   While the presence of thick tori predicted by unified schemes of AGN (see, e.g., Antonucci 1993) is relatively well established for powerful radio galaxies through detections of obscuring atomic or molecular gas, X-ray absorption or free-free absorption, the structure of the very inner regions of low-luminosity radio galaxies is still less defined.  The study of obscuration
toward the central regions of these galaxies (see e.g. Balmaverde, Capetti \&
Grandi 2006; Worrall et al.\ 2003, Chiaberge, Capetti \& Celotti 1999, Chiaberge
et al.\ 2002) has shown that the nuclear disks in these galaxies are
geometrically and optically thin. This suggests that the {\sl standard pc-scale
geometrically thick torus is not present} in these low-luminosity radio
galaxies.  

On the other hand, the presence of gas in the nuclear regions can also affect
the evolution of the radio sources (and in particular the radio jet) and the way
the radio sources grow. Interaction between powerful radio jets and their
environment is recognized to produce fast outflow that can be relevant in the
evolution of the host galaxy (see e.g. Morganti et al.\ 2005, Holt et al.\
2008). Entrainment of gas by the radio jet could also be crucial in slowing down
the jet from relativistic to sub-relativistic velocities (see Laing et al.
2006).
 
\begin{figure}  
\centerline{\psfig{figure=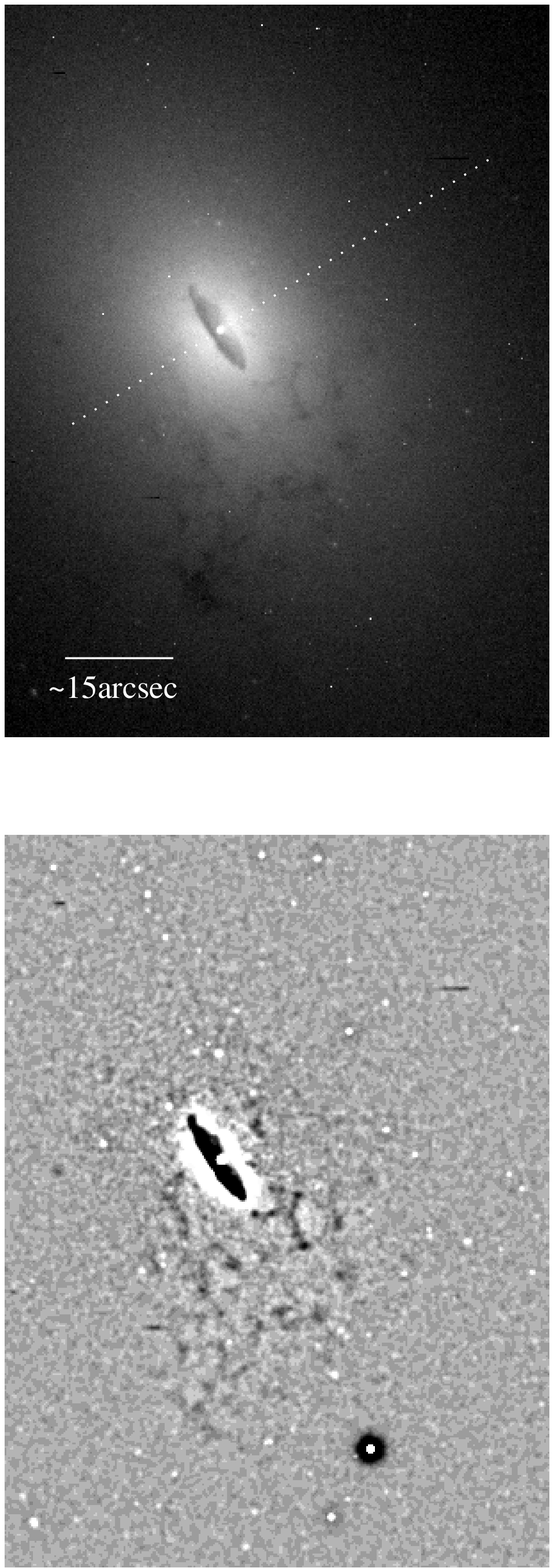,width=7.5cm}}
\caption{Top: Image of NGC~315 obtained from HST (Capetti et al. 2000). The optical core is clearly
visible as well as the dust disk and dust patches. The approximate direction of the radio jet is also indicated. Bottom: same image but highpass filtered and slightly smoothed afterwards to highlight  the dust patches around the nucleus.}
\label{fig:hst}
\end{figure}

\begin{table*}
\centering
\caption{Instrumental parameters of the \HI\ observations}
\label{tab1}
\begin{tabular}{lccc}
\hline \hline \\
Field Centre (J2000) & \multicolumn{2}{c}{RA = 0$^{\rm h}$ 57$^{\rm m}$ 48$^{\rm s}$;  Dec = 30$^\circ$ 21$^{\prime}$ 09$^{\prime\prime}$}   \\
\\
\hline 
\\
Instrument                & WSRT             & VLA A-array   & VLBI \\
Date of the observations  & 30Jun00, 05Sep01 &  04Dec00      & 25Nov01\\
Integration Time          &  11h, 10h        & 0.5h          & 4.5h\\
Synthesized beam (arcsec) & $35\times 18$ ($-5^\circ$) 
                          &  $1.2\times 1.0$ (26$^\circ$) 
                          & $9\times 3$mas (-20$^\circ$) \\
Number of channels        & 128              & 64            & 512 \\
Bandwidth (MHz)           & 10               & 6.5           & 8\\
Central frequency (MHz)   & 1397.4           & 1397.4        & 1396.18  \\
Velocity resolution (\kms) & 20              & 12            & 6.0 \\
rms noise in channel maps (mJy beam$^{-1}$)   & 0.21            & 0.86          & 0.7 \\
                &                 &         &   \\
\hline 
\hline \\
\end{tabular}
\end{table*}

Thus, it is clear that determining the gas distribution and kinematics in the
nuclear regions is of paramount importance for understanding radio galaxies in
general.  Among the different gas components, the neutral hydrogen is
particularly suited for this task.  Atomic neutral hydrogen has been observed in
absorption against the radio continuum in the central regions of a number of
radio galaxies (see e.g.\ Heckman et al.\  1983, Shostak et al.\  1983, van
Gorkom et al.\  1990, Morganti et al.\  2001, Vermeulen et al.\ 2003, Gupta et
al.\ 2006). In some cases, though clearly not all, this \HI\ absorption has been interpreted as due to gas distributed in
circumnuclear disks/tori although there are surprisingly few examples where the
spatially resolved signature of a rotating disk is actually observed.   However,
\HI\ absorption is not always found at the systemic velocity of the galaxy and,
as mentioned above, more sensitive observations and detailed studies show that
in many cases the kinematics of the gas can be disturbed and affected by the
presence of the radio jets (Morganti et al.\  2005 and references therein).
Understanding the effect of the radio jets  on this gas requires high
spatial resolution observations that are at present quite scarce.

Here we present the results from our detailed study of \HI\ in the radio galaxy
NGC~315.  This includes Westerbork Synthesis Radio Telescope (WSRT), Very Large
Array (VLA) and Global Very Long Baseline Interferometer (VLBI) observations. We
have looked at this galaxy as part of a larger project to study, at both low and
high resolution, the presence and the characteristics of the neutral hydrogen of
a representative sample of low-luminosity radio galaxies (see Morganti 2002 for some details).  The results of this
statistical study  will be presented in a forthcoming paper.

\subsection{Why NGC~315?}

NGC~315 is a bright elliptical galaxy.  An accurate systemic velocity of
4942$\pm$6 \kms\ ($z = 0.01648$\footnote{Throughout this paper we use a Hubble
constant $H_{\rm o}$= 70 km s$^{-1}$ Mpc$^{-1}$ and $\Omega_\Lambda=0.7$ and
$\Omega_{\rm M} = 0.3$. At the distance of NGC~315 this results in 1 arcsec =
0.335 kpc.}) has been derived from stellar absorption lines by Trager et al.\
(2000).  In the optical  band, NGC~315
shows a highly inclined, very regular, circum-nuclear disk, seen in absorption
in the HST image (see Fig.\ 1).  The dusty disk has a position angle of  about
40$^\circ$ and it extends to $r \sim 700$ pc with a mass of $\sim 1.9 \times 10^7 M_\odot$ (Forbes 1991) calculated from the S$_{\rm 100 \mu m}$ flux following Knapp, Bies \& van Gorkum (1990). At its centre there is an
unresolved optical compact core (Capetti et al.\ 2000, 2002).  A number of
discrete optical absorption clouds, west of the nucleus at distances ranging
from 3 to 7 arcsec, have been detected by Butcher et al.\ (1980). A mass limit
of $10^6 M_\odot$ has been derived for these clouds. The HST image shows
additional absorption clouds to the SW.  The dust disk is associated with a disk
of ionised gas which appears to be in ordered rotation (Noel-Storr et al.\
2003). CO emission was detected by Leon et al.\ (2003).  The inferred mass of
molecular hydrogen is $(3.0\pm 0.3) \times 10^8$ M$_\odot$.

NGC~315 hosts  a giant ($\sim$$1$ Mpc) radio source that is also known as
B2~0055+30 (Bridle et al. 1979, Mack et al.\  1998, Laing et al.\  2006 and references  therein).  The
 source has a radio power of log$P_{\rm 1.4~GHz} = 25.40$ W
Hz$^{-1}$ and  edge-darkened lobes (i.e.\ with  Fanaroff-Riley type I
structure, see Fig. 2). On  VLBI scales, the source shows  a core,  a bright jet
and a faint counter-jet (Cotton et al.\  1999).  The main jet is quite smooth. A
multi-epoch study shows evidence for moving features, indicating an
accelerating, mildly relativistic jet (Cotton et al.\  1999). The jets show deceleration at larger
distances and  detailed modeling indicates that the angle of
the jet to the line of sight is $\sim$$38^\circ$ (Canvin et al.\ 2005).

The nuclear X-ray emission of NGC~315 is seen through a moderate intrinsic
column density of $N_{\rm H} \sim 5\times 10^{21}$ cm$^{-2}$ (Worrall,
Birkinshaw \& Hardcastle 2003). Chandra typically finds column densities below
$10^{22}$  cm$^{-2}$ in FRI radio galaxy nuclei (e.g. Worrall, Birkinshaw \&
Hardcastle 2001), suggesting that the X-ray emission is associated with the
sub-kpc-scale radio jets.

\begin{figure*}  
\centerline{\psfig{figure=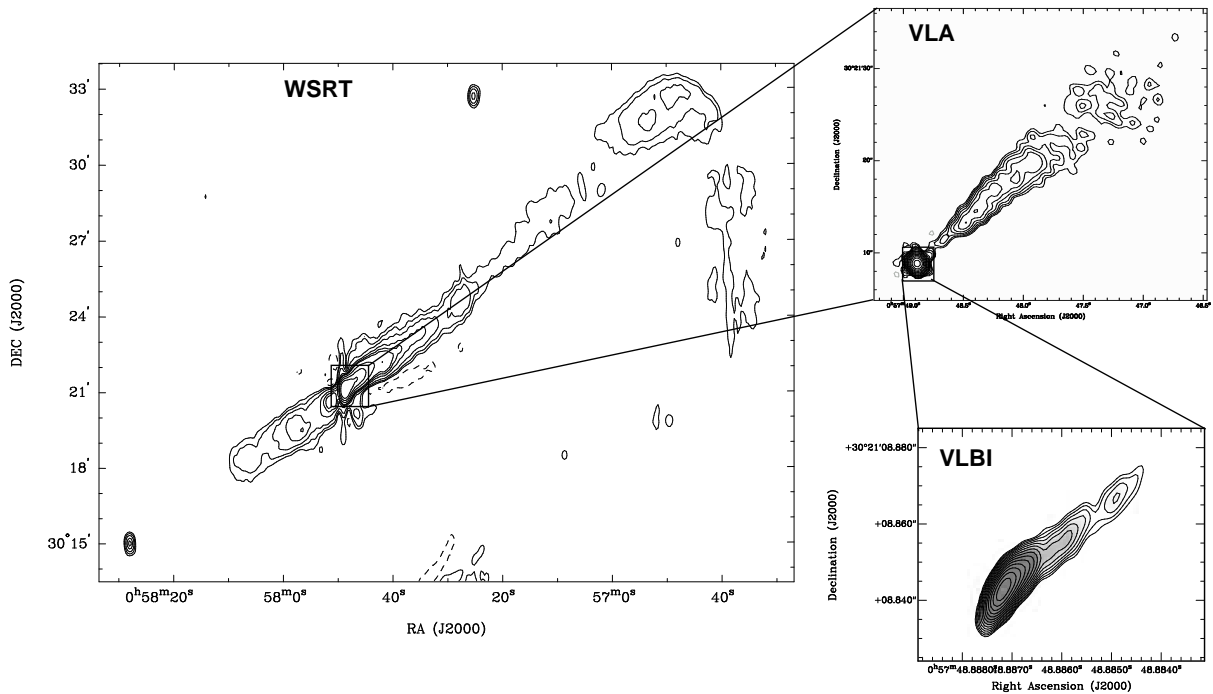,angle=0,width=16.0cm}
}
\caption{Radio continuum image of NGC~315 at different angular resolutions (see Table
1): WSRT (left), VLA (middle) and VLBI (right). The images were obtained from
the line-free channels, see text for details. Contour levels of the WSRT image
-1, 1, 2, 4, 8, 12, 16, 32, 64 and 128 mJy/beam.}
\label{fig:cont}
\end{figure*}

NGC  315 has been observed in \HI\ by different authors (Dressel et al.\ 1983,
Heckman et al.\  1983, Shostak et al.\  1983, Chamaraux et al.\  1987).  In
these studies, a narrow and highly redshifted ($\sim$$490$ \kms\ with respect to the
systemic velocity) \HI\ absorption component has been detected. This \HI\
absorption  splits in two components when  observed with high velocity resolution.
These two components are very narrow (velocity width $\sim$$2.5$ \kms) and very
deep ($\tau =$ 0.87 and 0.21) and separated by only $\sim 3$ \kms. A possible detection of an other 
broader \HI\ absorption closer to the systemic velocity was reported by Heckman
et al.\  (1983).

\section{WSRT and VLA observations and results}\label{h1obs}

The \HI\ observations of NGC~315 reported in this paper have been carried out
with various  angular resolutions. The different
observations have different purposes. In particular, we wanted to study in
detail not only the \HI\ absorption, but also investigate the possible presence
of \HI\ emission in and/or around this galaxy (Emonts et al.\ in prep.).

\subsection {Observations}

The low resolution WSRT observations were  done in order to investigate
the presence of neutral hydrogen, both in emission and absorption. The
relatively low spatial resolution of the WSRT is ideal when looking for low
surface brightness and extended structures.  The parameters of the observations
are summarized in Table~1.  3C~147 was used as flux and bandpass calibrator. The
data were calibrated and reduced using the MIRIAD package (Sault et al.\ 1995).
The continuum subtraction was done using a linear fit through the line-free
channels of each visibility record and subtracting this fit from all the
frequency channels ("UVLIN").   The spectral-line cube was obtained using robust
weighting set to 0.5 (Briggs 1995), i.e.\, intermediate between natural
and uniform weighting.    The data were Hanning smoothed to suppress the Gibbs
ripples produced by the strong narrow absorption present in this galaxy. The
resolution is $35\times 18$ arcsec (p.a.\ $-5^\circ$) and the r.m.s.\ noise
level is $\sim 0.21$ mJy beam$^{-1}$. In these observations we detected two \HI\
absorption systems, as well as  \HI\ emission from  NGC 315 and from five
neighbouring galaxies. A continum image was obtained from the line-free channels
and  is shown for reference in Fig.~2.


VLA observations of NGC~315 were obtained as part of a larger project aimed at
studying the possible relation between the presence of \HI\ absorption, the
characteristics of the nuclear dust disks (seen by HST, Capetti et al.\ 2000)
and the presence of optical cores (see Morganti et al.\ 2002 for preliminary
results). The VLA observations have been carried out using the A-array to profit
from the highest possible resolution with this telescope.  The parameters are
summarized in Table~1.  The data reduction was done in a similar way as for the
WSRT observations, again using the MIRIAD package.  The continuum (from the
line-free channels) and the line cube were produced with uniform weighting.

\subsection{The two \HI\ absorption systems}

The first  result of these observations is that {\sl two} \HI\ absorbing systems
are detected in NGC~315. Apart from the  very narrow and redshifted
component (see Fig.\ 3), we also clearly detect the broad component that Heckman
et al.\ (1983) reported as a probable detection. It is worth mentioning that
there are only two other cases known of radio galaxies where two clearly
separated \HI\ absorbing systems have been observed:  NGC~1275 (3C~84, see van
Gorkom et al.\ 1989 and references therein) and 4C~31.04 (Mirabel 1990, Conway 1996).

The two components in NGC~315 are detected in all observations, also at the VLBI
scale (see Sec.\ 6).  Fig.~3 shows the profiles of the two components.  As
already known (Shostak et al.\  1983, Dressel et al.\  1983), we find that the
narrow \HI\ absorption is highly redshifted (by $\sim$$490$ \kms) compared to the
systemic velocity. The narrow absorption is known to have a FWHM of only 2.5
\kms\ (Dressel et al.\ 1983).   Note that the observed width of the narrow, redshifted \HI\ absorption is limited by the low velocity resolution (and Hanning smoothing) of our data, therefore appearing much larger than then $\sim 2.5$ \kms\ detected with high velocity resolution by Arecibo (Dressel et al. 1983).Therefore, our low velocity resolution
observations do not provide any new insight on this component.  The integrated
absorbed flux that we derive (977 mJy \kms) is also consistent with that
obtained from Arecibo  (998 mJy \kms).  

The broad \HI\ absorption, detected in both the WSRT and the VLA observations,
has a full-width zero intensity (FWZI) in velocity of $\sim$$150$ \kms\  (FWHM $\sim 80$ \kms) and it is centered on $\sim 5020$
\kms. Although the broader component is located closer to the systemic velocity
than the narrow component, it is nevertheless redshifted by about 80 \kms. For
this galaxy an accurate systemic velocity derived from  stellar absorption lines
is available ($V_{\rm sys} = 4942 \pm 6$ \kms; Trager et al.\  2000). This estimate of the systemic velocity is  not affected by the
disturbed kinematics that the ionised gas may have, thus the
offset in velocity is significant.

The characteristics of both \HI\ absorption components, when accounting for
the different velocity  resolution, are not different between the
WSRT and the VLA data (see Fig.\ 3), indicating that the structures that produce
the absorption have a size smaller than $\sim 1$ arcsec (0.5 kpc) and therefore
the study of their morphology requires VLBI observations (see Sec.\ 6).

The peak of the continuum emission is 531 mJy in the WSRT data, while the peak
absorption of the broad component is $\sim$$-4.8$ mJy and the integrated
absorbed flux that we derive is $\sim 353$ mJy \kms.  The resulting optical
depth is $\sim$$0.009$ and the corresponding column density of $2.5 \times
10^{18}$ $T_{\rm spin}/f$ cm$^{-2}$  where $f$ is the covering factor.  Similar values for the optical depth are
derived from the VLA data.

\begin{figure} \centerline{\psfig{figure=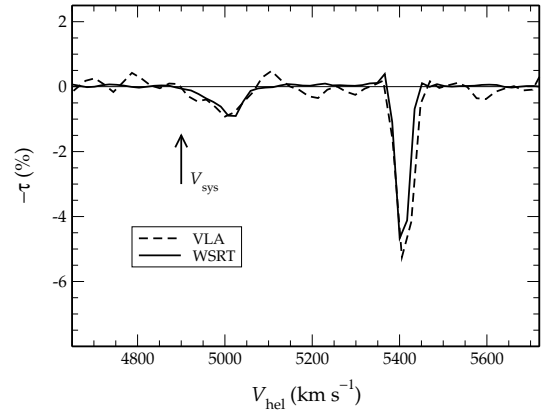,angle=0,width=7cm}}
\caption{\HI\ profile (in optical depth) obtained from the 
 WSRT (solid line) and the VLA (dashed) data.  The two \HI\ absorption systems are
clearly visible.  The systemic velocity from Trager et al.\ (2000)  is also
indicated. The differences in the profile, in particular for the narrow component, is likely due to the different velocity resolution of the two observations. The bump in the WSRT profile at velocities just below those of the narrow absorption, corresponds to the \HI\ emission detected in NGC~315.} \end{figure}

\subsection{\HI\ emission and the environment}\label{emission}

One of the aims of the low-resolution WSRT observations was to look for  \HI\ in
emission in and around NGC~315. We do indeed find  such emission in NGC 315
(Fig.\ 4 right), at the same location as the HST images show faint optical
absorption features. The emission is detected at velocities from $V_{\rm hel} =
5340$ \kms\ up to that of the narrow absorption. The \HI\ mass of the emission
is $6.8 \times 10^7\ M_\odot$ and the peak column density is $1.0\times
10^{20}$ cm$^{-2}$.   NGC~315 is, therefore, similar to other FRI radio galaxies
(Morganti et al.\ 2009, Emonts et al.\ in prep). Radio galaxies of this type do
not appear to have large amounts of \HI\ associated with the host galaxy (Emonts
2006, Emonts et al.\  2006). Large amounts of neutral hydrogen have so far been
detected  only near {\sl compact} radio galaxies (Emonts et al.\ 2006). In these
sources, the gas is distributed in very extended disks/rings
with large \HI\ mass ($\gta 10^9$ $M_{\odot}$), possibly indicating that a
major merger has occurred in the not too distant past.  This lack of  such
large, disk-like \HI\ structures in extended FRI sources can be seen as another
indication that triggering of the AGN does not happen through {\sl major}
mergers.

\begin{table}
\centering
\caption{\HI\ emission from other galaxies in the field of NGC~315. The last
column (sep) gives the distance (in kpc) of these galaxies from NGC~315.}
\label{tab2}
\begin{tabular}{lccccc}
\hline\hline\\
Name & \multicolumn{2}{c}{Position (J2000)} & $V_{\rm sys}$ & $M_{\rm HI}$ & sep \\
     &    RA & Dec     &   \kms  & $10^7\ M_\odot$ & kpc  \\
\hline \\
 J05730+3011     & $00^{\rm h}57^{\rm m}30^{\rm s}$ & $30^\circ 11^\prime 05^{\prime\prime}$     &    4888  & 66.7 & 214    \\
 J05819+3016    & $00^{\rm h}58^{\rm m}19^{\rm s}$ & $30^\circ 16^\prime 05^{\prime\prime}$     &    5280  & 4.0  & 164    \\
 J05755+3024     & $00^{\rm h}57^{\rm m}55^{\rm s}$ & $30^\circ 24^\prime 32^{\prime\prime}$     &    4208  & 38.2 & 75    \\
 J05642+3027    & $00^{\rm h}56^{\rm m}42^{\rm s}$ & $30^\circ 27^\prime 13^{\prime\prime}$     &    5007  & 77.8 & 308    \\
 J05700+3027    & $00^{\rm h}56^{\rm m}59^{\rm s}$ & $30^\circ 27^\prime 15^{\prime\prime}$     &    5246  & 83.4 & 246    \\
     &          &           \\
\hline\hline \\
\end{tabular}
\end{table}

Apart from the emission in NGC 315, a number of neighbouring galaxies are
detected in \HI. NGC 315 is part of the Zwicky cluster 0107.5+3212 (Zwicky et
al.\ 1961) which is located in the Perseus-Pisces filament.  Garcia (1993) lists 
18 galaxies as part of this group. We detect \HI\ emission in 5 galaxies  within
a  few hundred \kms\ from the redshift of NGC~315.  The objects and their
characteristics are listed in Table 3. All galaxies have an optical counterpart
(two of them are 2MASS sources) and are not part of the list of Garcia (1993).

In Fig.\ 4, we give the total \HI\ intensity image, showing the 5  companions. The environment of NGC 315 turns out to be quite gas rich, the total amount of \HI\ within 400 kpc of NGC 315 is $2.7\times 10^9\ M_\odot$.  It is clear that NGC 315 is the dominant galaxy of the group. In such an environment it is not unlikely that a small, gas-rich companion has fallen into NGC 315.

\begin{figure*}
\centerline{\psfig{figure=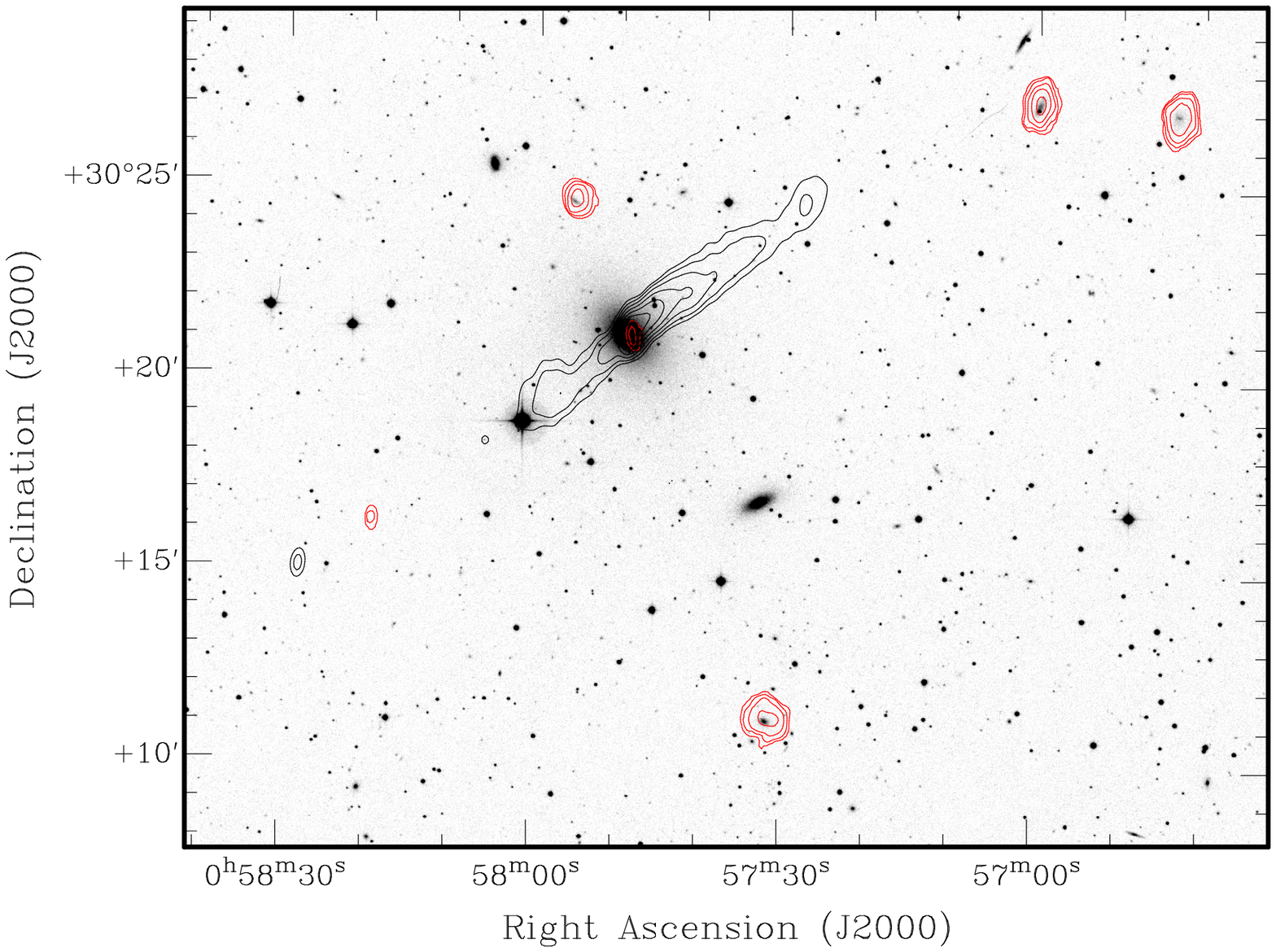,angle=0,height=7cm}
\psfig{figure=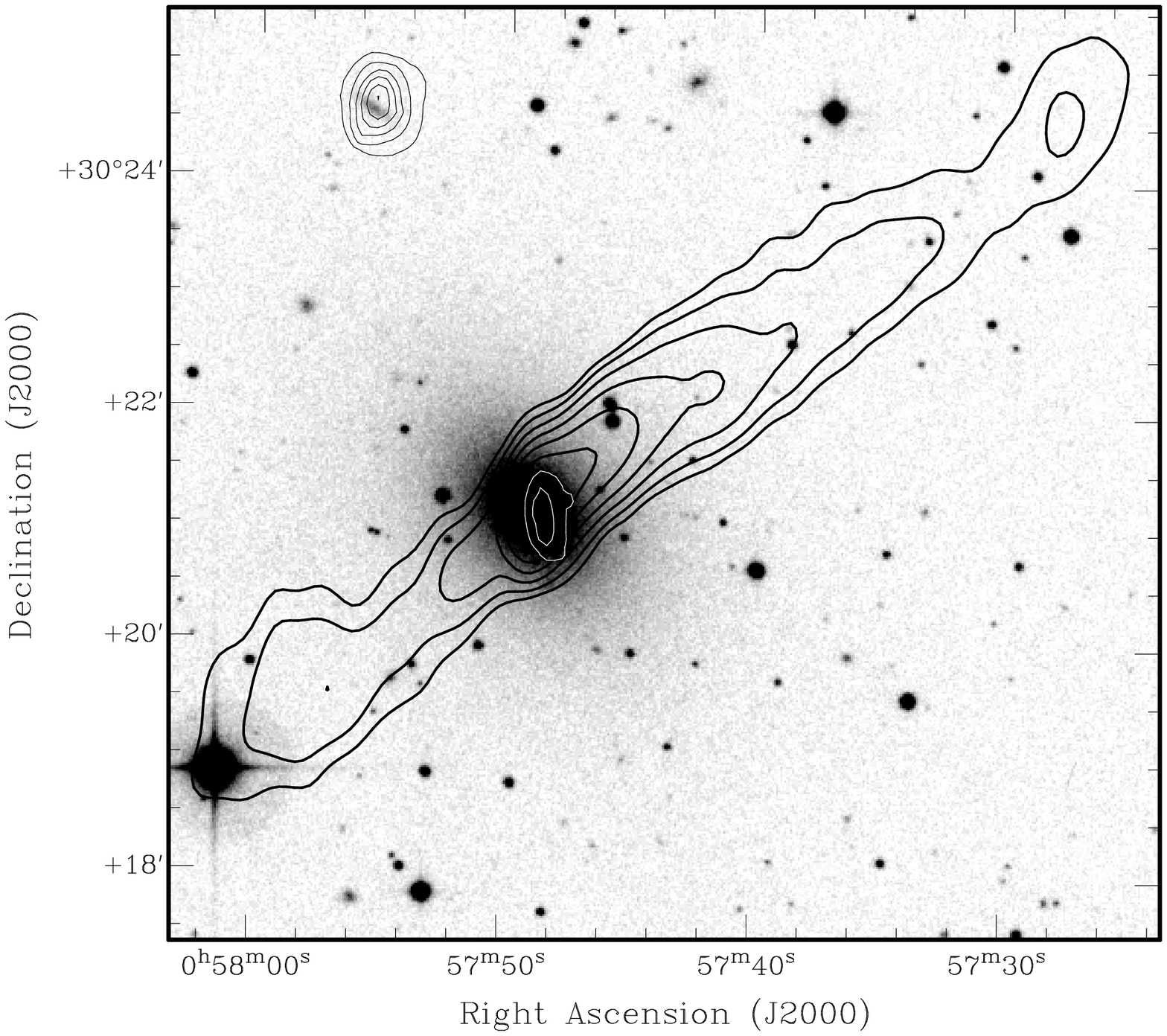,height=7cm}}
\caption{{\sl left:} Contours (red) of the integrated \HI\ emission derived from the WSRT observations, on top of an optical image obtained from the DSS. Contour levels are 4, 8, 16 and $32\times10^{19}$ cm$^{-2}$. Also given are the contours (black) representing the continuum image derived from the same data. Contour levels of the continuum are 5, 10, 20, 40, ... mJy beam$^{-1}$. {\sl Right:} Detail of the figure on the left, showing the \HI\ emission detected in NGC 315 and one nearby companion (thin contours) as well as
 the continuum (thick contours). Same contour levels as on the left.
}
\label{fig:mom1}
\end{figure*}

\begin{figure*}  
\centerline{{\psfig{figure=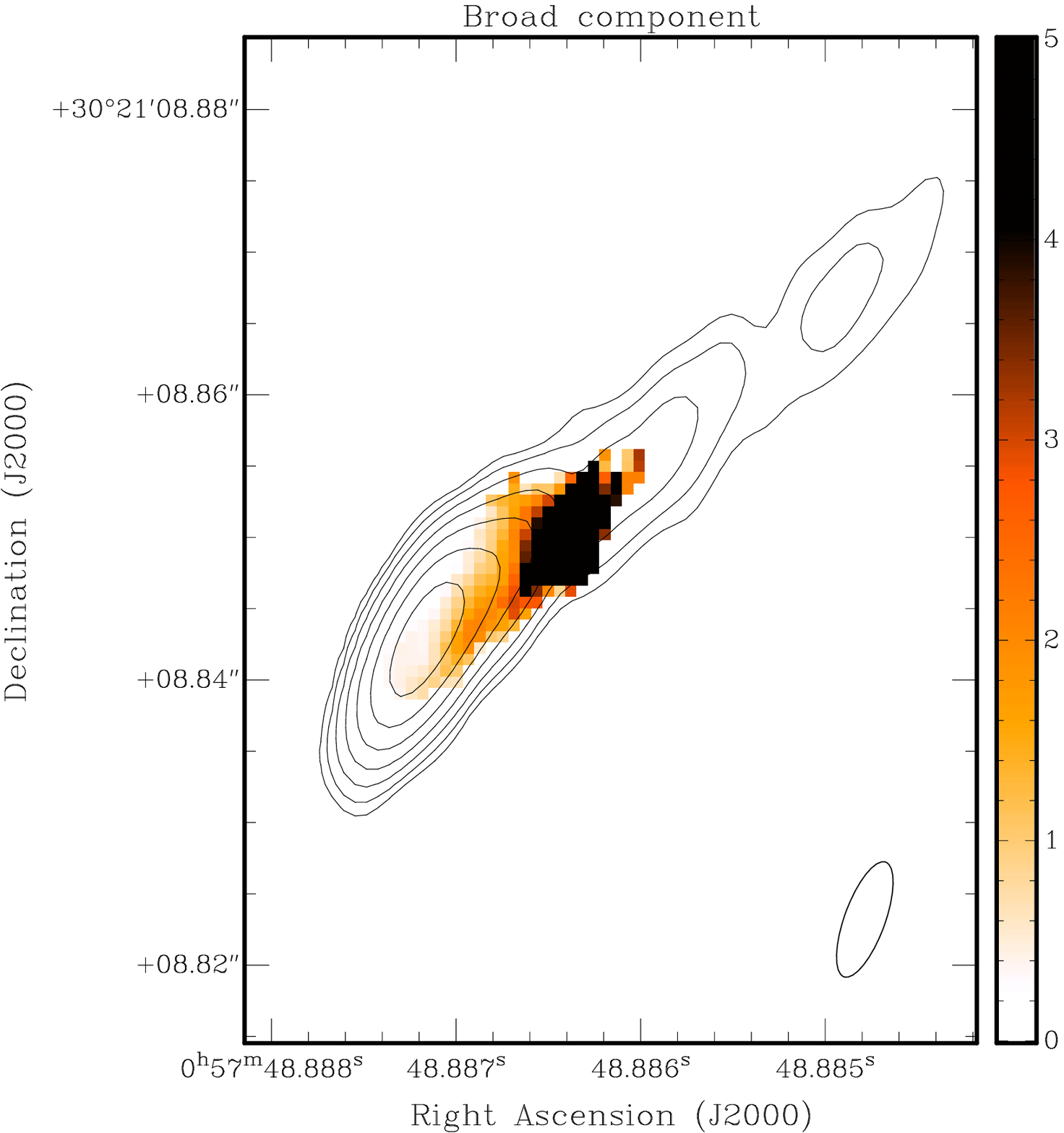,angle=0,width=8cm} 
\psfig{figure=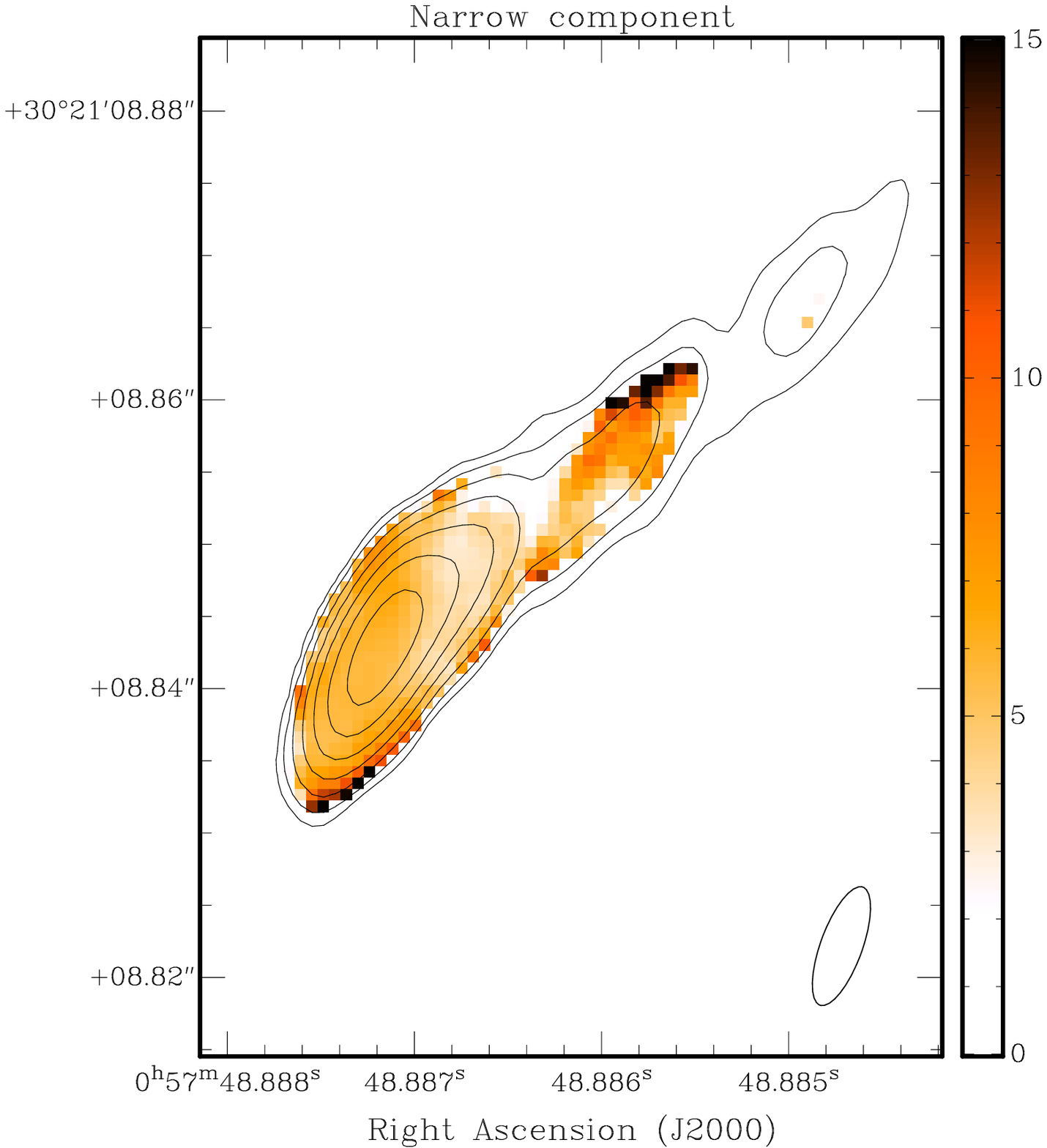,angle=0,width=8cm}
            }}
\caption{Continuum (contours) and column density (grayscale) of the
broad (left) and narrow (right) \HI\ absorption as derived from the VLBI observations. The intensity scales given on the right side of each panel are in units of $10^{20}$ cm$^{-2}$. The contours of the continuum are 2, 4, 8, 16, ... mJy beam$^{-1}$. 
}
\end{figure*}

\section{VLBI observations}

Earlier VLBI \HI\ observations of NGC~315 have been  presented by Peck (1999).
However, in these observations the  observing band was centered on the velocity
of the narrow \HI\ absorption, as the presence of a broad absorption component
was not yet established at the time.  The new VLBI observations presented here
use a broader bandwidth and a central frequency that allows us to detect  both \HI\
components.

NGC~315 was observed  with the Global VLBI Network in November 2001. We used a
bandwidth of 8 MHz in 512 channels. The velocity resolution, after Hanning
smoothing, is $\sim$$6$ \kms.  The rms noise is $\sim 0.26$ mJy beam$^{-1}$ in
the continuum image and $\sim 0.9$ mJy beam$^{-1}$  in the line channels.

The continuum image obtained from the line-free channels (beam $9\times 3$ mas
pa = $-20^{\circ}$ using natural weighting) is shown in Fig.~2. A strong core
and a jet  are detected, consistent with  previous observations (see e.g.\ the
detailed observations presented in Cotton et al.\ 1999).  The position angle of
the jet is about 43$^\circ$, consistent with that of the large-scale jet and
perpendicular to dust disk observed by HST (see Fig.~1). The peak of the
continuum is about 190 mJy beam$^{-1}$.  In the following sections we present
the results obtained from the VLBI line data.

\section{The \HI\ absorption at pc-scale and the origin of the atomic neutral hydrogen}

Both the narrow and the broad \HI\ absorption systems are detected and spatially
resolved by the VLBI observations. As we discuss below, both the kinematics
and the column density distribution are quite different for the two components,
suggesting that the two absorption systems have very distinct origins.

\subsection{The broad \HI\ absorption}

An interesting but puzzling result from the VLBI observations is the morphology and
kinematics of the broad \HI\ absorption (Figs 5 left and 6). Some absorption is
detected against the core, but there is a strong gradient in column density, the
column density being much higher further down the jet. Together with the density
gradient, there is also a velocity gradient along the jet, with the gas further
away from the nucleus having a larger redshift. The derived column density has a
peak of $\sim 1 \times 10^{19} T_{\rm spin}$ cm$^{-2}$/K (where the filling factor $f$ has been taken as 1 given that the absorption is resolved), while the column density
in front of the nucleus is about a factor 10 lower. Further down the jet, no
absorption is detected, but this is most likely due to the limited sensitivity
of the VLBI  observations. Indeed,  about 30\% of the flux detected for the
broad absorption in the WSRT and VLA observations is missing in the VLBI
observations.

One can attempt to make a rough estimate of the mass of the \HI\ producing  the
broader absorption.  Taking an average column density of $2 \times 10^{20}$ cm$^{-2}$  (for a canonical $T_{\rm spin} =100$ K) and
an area of $10 \times 3$ parsec (corresponding to $30 \times 10$ mas), the mass enclosed
would be about 100 $M_{\odot}$. This number is most likely a lower limit 
because of a number of uncertain parameters.  Most uncertain is perhaps the size
of the gas cloud:  part of the flux is missing and, therefore, the absorption is
likely to be more extended than  can be recovered from VLBI observations. 
In addition, of course, the \HI\ detected is only that part of the gas cloud
that  is in front of the jet. Moreover,  the assumed value for the $T_{\rm
spin}$ is uncertain. The gas producing the \HI\ absorption  may be located
very close to the nucleus (see next section) and therefore the assumption of
$T_{\rm spin}=100$ K could be unrealistic. The $T_{\rm spin}$ might well  be at
least a factor 10 higher (see also results on PKS~1549--79 by Holt et al.\ 2006).  X-ray observations suggest  that the AGN is seen through a medium with a column density of $5\times10^{21}$ cm$^{-2}$ (Worrall et al.\ 2003), indeed suggesting that the spin temperature is substantially higher than 100 K.
Therefore, 1000 $M_{\odot}$ is a more likely lower limit to the \HI\ mass.
In this context, it is quite conceivable that the observed gradient in column density more reflects a gradient in $T_{\rm spin}$  and not a gradient in the actual density.

\begin{figure}  
\centerline{\psfig{figure=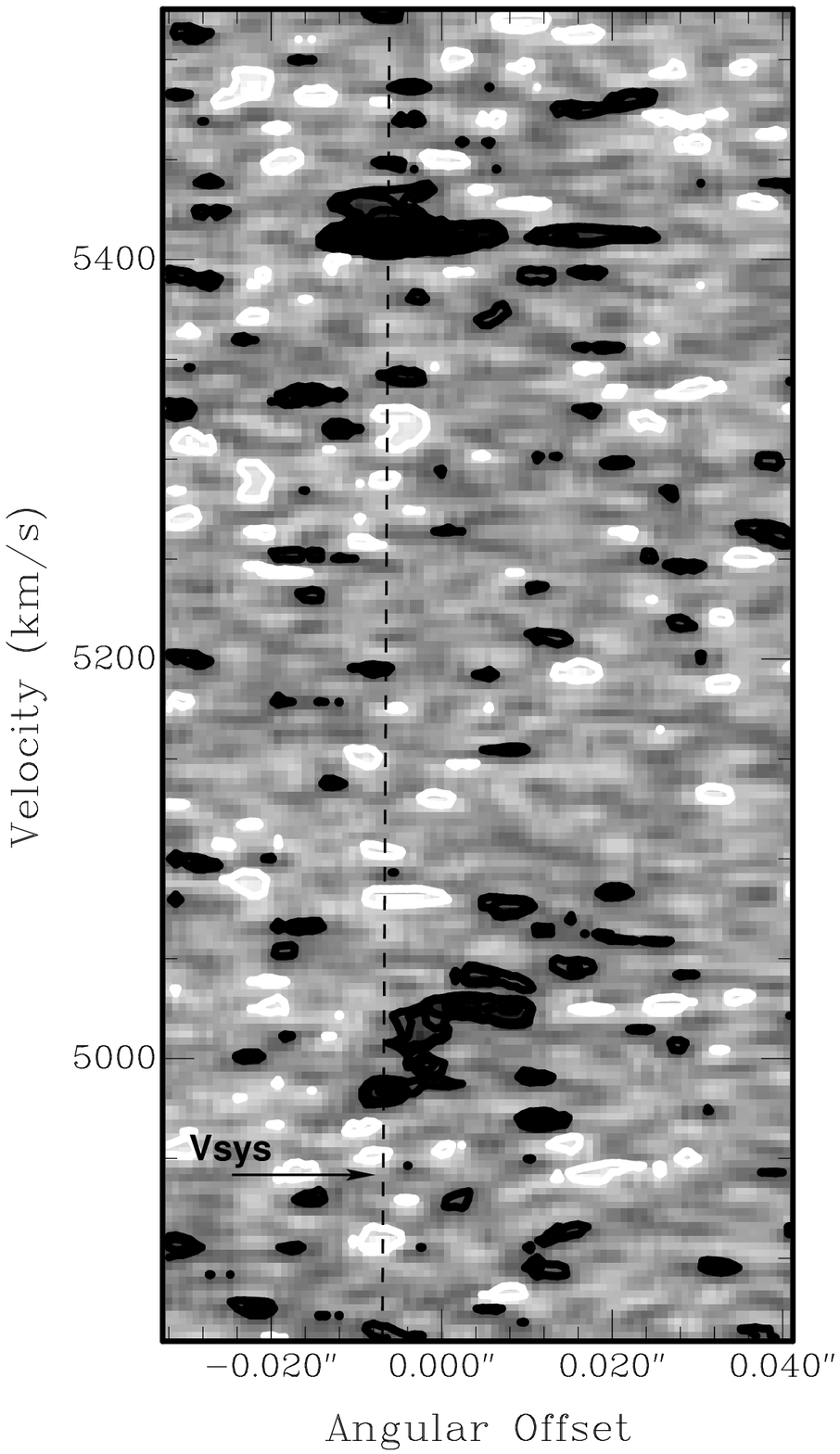,angle=0,width=10cm}}
\caption{Position-velocity plot taken along the radio jet (p.a. $\sim 130^{\circ}$). The plot shows  both
absorption features, centered around 5000 and 5400 \kms\ respectively. The velocity gradient in the \HI\ broader component is visible. The contour levels are -0.0019, -0.0025, -0.0038, -0.01, -0.03, 0.0019 mJy beam$^{-1}$. Negative contours (representing the absorption) are in black. The position of the core and the systemic velocity are indicated}
\end{figure}

\subsection{Nature of the cloud producing the broad \HI\ absorption}\label{broad}

The strong gradients in the \HI\ column density  and velocity  of the broad absorption suggest that the gas is being influenced by the presence of the AGN, indicating that this gas is physically close to the core of NGC315.  Moreover, as has been noted about broad \HI\ absorption in other
objects (see for example  the case of Cygnus A described in Conway \& Blanco 1995),  a
major problem of  {\sl broad} absorption being due to a chance superposition of
a foreground cloud (e.g.\ similar to the dust patches  SW of the core seen in
the optical)  is the  large velocity width and gradient of the  broad \HI\
absorption. For clouds in the ISM of galaxies, even if one considers clouds that
are "wreckage" of some accretion event, one would  expect to observe  velocity
gradients much smaller than  the 100 \kms\ over only 7 pc observed for the broad
absorption.

However, although it is likely that the broad absorption occurs close to the AGN, the nature of the gas resulting in the broad absorption is not clear. One interpretation which may fit the data is that of a circumnuclear torus.  A configuration in which we are looking through a section of a thick rotating torus comprised mainly of atomic gas would result in a broad line width.  In cases where the radio source axis is in the plane of the sky and our line of sight passes through the midplane of the torus, even broader lines have been found, and in almost all cases detected to date, absorption features attributable to a circumnuclear torus are redshifted with respect to the systemic velocity of the host galaxy (e.g. PKS 2322-123, FWHM$\sim$700 km s$^{-1}$ redshifted by $\sim$220 km s$^{-1}$; Taylor et al 1999). This is probably due to streaming motions within the structure toward the accretion disk and central supermassive black hole.  

In a source which has a jet axis close to the plane of the sky, we would expect to see absorption from a torus toward the inner jets on both sides of the core (e.g. Peck, Taylor \& Conway 1999), with a smaller optical depth possible toward the core due to much higher spin temperatures near the AGN, as described above.  However in a radio source which is oriented more toward our line of sight, such as NGC~315, the counterjet is Doppler-dimmed, and the plane of the disk is not aligned with our vantage point.  Thus we do not necessarily expect to have the sensitivity to detect absorption toward the inner counterjet, as the continuum source is very faint, and we are looking along a line of sight from the top through the middle of the torus, where much of the gas may lie in the region of higher spin temperature and thus yield a lower optical depth.  

Where the circumnuclear torus model struggles in this source, however, is in explaining the compact region of higher column density shown in Fig 5.  Some circmnuclear tori have been shown to be clumpy on similar scales (Peck \& Taylor 2001), but not with this range of density.  Additionally, it would be hard to explain the velocity gradient along the jet unless this might be due to the gas above the midplane of the torus streaming down and inward toward the central mass.  Such a scenario would result in an inward component of the velocity vector which would point more directly away from our vantage point as one went higher in scale height, yielding a larger redshift.  The gas further down the jet would also have a component of inward streaming motion in addition to rotation, but given the orientation of the torus, these motions would not be directly away from us, and thus would not give rise to as large a redshift. Alternatively, one might explain both the compact region of high column density and the cospatial region of higher redshift with a discrete cloud falling toward a circumnuclear structure, but we do not have the spatial or velocity resolution to justify this level of detail in the model.  Thus at present, we cannot rule out the possibility of a circumnuclear toroidal structure, but more corroborating data would be required before we could adopt this model.


Another possibility is that the  broad absorption is due to gas entrained in the
jet. NGC 315 does contain a number of small gas/dust clouds (see below),  so it
is, in principle, possible that one of these is now interacting with the jet. A
structure of (molecular) gas aligned with radio jet has been found in M51
(Matsushida, Muller, Lin 2007). Also in that object, a velocity gradient with 
blueshifted velocities  closer to the nucleus has been observed. The authors
argue that  the observed structure and velocity gradient are due to molecular gas
entrained by the radio jet. Naively, one would however expect that if gas is
entrained in a jet, the velocities  would appear more blueshifted (i.e. coming toward the observer dragged by the jet) as one goes
along the jet, opposite to these observations. We note in passing that  the broad absorption is
redshifted with respect to the systemic velocity, but that it is  blueshifted
relative to the narrow absorption. If the velocity of the narrow absorption is
representative for small clouds falling into NGC 315, the jet would
then be pushing the gas out with  (projected) blueshifts of  about 400 \kms. One
complication is that  it would be have to be a coincidence that the velocity of
the outflowing gas ends up to be close to the systemic velocity of the galaxy.
It would also still not explain the velocity gradient along the jet. 


Finally, let us consider the possibility that the  broad absorption is somehow 
related to the process of providing (directly or indirectly) material for 
fuelling  the activity in the nucleus, i.e. gas (not necessarily part of a circum-nuclear structure) that is falling into the nucleus. One can estimate, albeit with
considerable uncertainty, the mass accretion rate that would be associated to
this process, simply from the \HI\ mass and the timescale for falling in. 
For these calculations it is, however, important to keep in mind (as described in details above) that the \HI\ detected in absorption may be just a small fraction of the \HI\ present around the nucleus. This is because we are limited to the gas {\sl in front} of the radio continuum. This means that the numbers derived here (in particular the accretion rate) can be quite conservative lower limits.
Taking 50 \kms\ as the infalling velocity of the gas, the timescale would be
$\sim 7 \times 10^5$ yrs. This would give accretion rates of at least  $10^{-4}$
$M_\odot$ yr$^{-1}$ and more likely $10^{-3}$ $M_\odot$ yr$^{-1}$, if we adopt
the mass of the gas estimated above with the assumption for $T_{\rm spin} =
3000$ K.  An accretion rate of $10^{-3}$ $M_\odot$ yr$^{-1}$ is very similar to
those derived for other X-ray luminous elliptical galaxies, based on  Chandra
observations  (Allen et al.\ 2006). On the other hand, using the luminosity of
the radio core, one can estimate the power of the jet and from that the
accretion rate (see Balmaverde et al.\ 2008). 
Starting from a radio core flux of 210 mJy, we find $P_{\rm core} \sim 10^{30}$ erg s$^{-1}$ Hz$^{-1}$.
Using the relationship between core luminosity and jet power (Heinz et al.
2007) this corresponds to  log $P_{jet}
\sim 44.15$ erg s$^{-1}$ Hz$^{-1}$. This can be converted in an accretion power of  log $P_{accretion} \sim  45.8$ erg
s$^{-1}$ Hz$^{-1}$ (Balmaverde et al. 2008) implying 
a mass accretion rate $\dot M_\odot \sim 0.1$
M$_\odot$ y$^{-1}$.  This accretion rate is quite high when compared to e.g.\ the sample in Allen et
al.\ (2006). This high feeding rate is consistent with the high core power in NGC~315 relative to
the galaxies in that sample. This would suggest that in the case of NGC~315,
the hot gas dominates the accretion, unless the physical size of the cloud
responsible for the broad absorption is much larger (a factor 100) than what we
have assumed.

It is worth mentioning that in the case of the powerful radio galaxy Cygnus A, a cloud of molecular gas has been detected along the radio jet. This cloud (that has a velocity redshifted compared to the systemic velocity of the host galaxy) has been interpreted as falling into the nucleus and perhaps  connected  to  fueling  the AGN (Bellamy  \& Tadhunter 2004). These authors argue that the connection may also be indirect, with the process of capturing the cloud and the subsequent settling into the circum-nuclear disk leading to radial motions in the disk and increased fueling rate. The tentative detection of an \HI\ counterpart to this cloud in Cygnus A (Morganti et al.\ in prep.)  makes the similarity to  NGC 315 more evident.

\subsection{The narrow \HI\ absorption}

Like the broad absorption,   the narrow \HI\ absorption is also spatially
resolved in the VLBI observations.  This absorption component covers about 20
mas of the source, $\sim$$9$ pc, from the core to the first part of the jet (see
Fig.\ 5 right). The narrow absorption is different from the broad one in all
respects.   The column density distribution is much more uniform, the velocity
is much narrower and  no velocity gradient is observed. The fact that, contrary
to what is seen for the broad absorption, the properties of the narrow
absorption do not show any relation with the location of the AGN, or gradient
along  the jet, suggests that the cloud producing the narrow absorption is not
physically close to the AGN. The very narrow velocity width also  supports this
(see also Dressel 1983). Moreover,  the narrow absorption is close  in
 in velocity (and perhaps also in space, i.e. at large radii) to the \HI\ emission detected in the WSRT observation.
This suggests that  the narrow absorption is due to a cloud similar to those
seen in optical images as absorbing clouds out to several kpc SW of the nucleus
(Fig.\ 1). Using HST data, de Ruiter et al.\ (2002) derive a typical gas column
density of $2.7 \times 10^{20}$ cm$^{-2}$ for the gas clouds seen in NGC 315, a
value very similar to that found in our radio observations  for the \HI\ in emission.
We obtain the same column density for the narrow absorption if we assume 
$T_{\rm spin} = 100$ K. This could further support the idea that the gas is at  larger radius since such value for $T_{\rm spin}$ is characteristic for the ISM not affected by the active nucleus. If these clouds are the result of a
small accretion event, one would indeed expect  a redshifted velocity (Dressel
1983). Our low-resolution observations have shown that the environment of NGC
315 contains several small, gas-rich galaxies, so accretion of a small companion
by NGC 315 is likely to have happened.

There are only two other cases known of radio
galaxies where {\sl two} such distinct \HI\ absorbing systems have been observed.  NGC~1275 (3C~84, van Gorkom et al.\ 1989 and  references therein) and 4C~31.04 (Mirabel 1990) In both cases, a redshifted narrow component has been observed
in addition to a broad closer to the systemic.  For both objects, the redshifted absorption has been explained as being due to material falling into the main galaxy and it appears that the same explanation also applies to NGC 315.

\section{Summary}

We have studied the properties of the \HI\  in the radio galaxy NGC 315. Two \HI\ absorption components are present, a broad one (FWZI $\sim$150 \kms)   redshifted ($\sim$80 \kms) with respect to the systemic velocity, and a very narrow component (
FWZI  $\sim 8$ \kms). 
Interestingly, the two absorption components have very different properties.
We also detect  \HI\ emission in NGC 315, a few kpc SW of the nucleus. This emission is likely associated with absorption patches that are observed in optical images. 

The broad absorption shows a strong gradient in 
column density (or spin temperature) along the jet, with the highest densities (or lowest spin temperatures) furthest away from the AGN. It also shows a strong velocity gradient (more than 100 \kms\ over 10 pc) with the more redshifted velocities away from
 the AGN. The properties of this broad component strongly suggest that the gas producing the absorption is physically close to the AGN. 
 The possibility that it is entrained by the radio jet (and partly responsible of the deceleration of the jet) appears unlikely because of the redshifted velocities of the gas.
Gas located in a thick circum-nuclear toroidal structure, with orientation 
similar to the dusty, circumnuclear disk observed with HST,  cannot be completely ruled out although it appears  difficult to reconcile with the observed morphology of the absorption and it would require inward streaming motion in addition to rotation. 
The scenario we favour,  is that the gas producing the broad absorption could be (directly or indirectly) connected with the
fuelling of the AGN, i.e. gas  that is falling into the nucleus. If this is the case, the accretion rate derived is similar
(considering all uncertainties) to that found for other X-ray luminous
elliptical galaxies, although lower than that derived from the radio core
luminosity for NGC~315.   

 On the other hand, the properties of the narrow absorption are very uniform. Moreover, it tightly connects, in space and in velocity, with the \HI\ emission in NGC 315. Most likely, the cloud responsible for the narrow absorption is quite far from the AGN and is likely due to material falling into NGC 315. Five nearby, small gas-rich companions are also detected in \HI. This implies that the environment of NGC 315 is quite gas rich and that accretion of gas from the environment is quite likely.

\begin{acknowledgements} We would like to acknowledge Natascha Boric. Part of
this work was done during her ASTRON/JIVE Summer Student project 2002.  We would
like to thanks Jes\'us Gonz\'alez for further inspecting the optical spectral of
NGC~315 and provide us the most accurate value of the systemic velocity. The
Westerbork Synthesis Radio telescope is operated by the ASTRON (Netherlands
Institute for Radio Astronomy) with support of the Netherlands Foundation for
Scientific Research (NWO). The National Radio Astronomy Observatory is a
facility of the National Science Foundation operated under cooperative agreement
by Associated Universities, Inc. This research has made use of the NASA
Extragalactic Database (NED), whose contributions to this paper are gratefully
acknowledged.  The Digitized Sky Survey was produced at the Space Telescope
Science Institute under US Government grant NAG W-2166. The European VLBI
Network is a joint facility of European, Chinese, South Africa and other radio
institutes funded by their national research councils.

\end{acknowledgements}


\begin{thebibliography}{}
\bibitem[]{}Antonucci, R. 1993, ARA\&A, 31, 473
\bibitem[]{}Allen S.W., Dunn R.J.H., Fabian A.C., Taylor G.B., Reynolds C.S. 2006, MNRAS 372, 21

\bibitem[]{} Balmaverde, B.; Baldi, R. D.; Capetti, A., 2008 A\&A 486, 119
\bibitem[]{} Balmaverde, B.; Capetti, A.; Grandi, P. 2006, A\&A 451, 35
\bibitem[]{} Bellamy M.J., Tadhunter C.N. 2004 MNRAS 353, 105
\bibitem[]{} Best P.N. et al. 2005, MNRAS 362, 25
\bibitem[]{} Best P.N. et al. 2007, New Astronomy Review 51, 168 
\bibitem[]{} Bridle A.H., Davis M.M., Fomalont E.B., Willis A.G., Strom R.G. 1979, ApJ 228, L9
\bibitem[]{} Briggs, D. 1995, Ph.D. thesis, New Mexico Inst. Mining Tech. 
\bibitem[]{} Butcher H.R., van Breugel W., Miley G., 1980, ApJ 235, 749
\bibitem[]{} Canvin, J. R.; Laing, R. A.; Bridle, A. H.; Cotton, W. D. 2005, MNRAS, 363, 1223
\bibitem[]{}Capetti A., de Ruiter H.R., Fanti R., Morganti R., Parma P., Ulrich  
M.-H. 2000, A\&A 362, 871;
\bibitem[]{}Capetti A., Celotti, A.; Chiaberge, M.; de Ruiter, H. R.; Fanti, R.; Morganti, R.; Parma, P. 2002, A\&A 383, 104

\bibitem[]{} Chamaraux P., Balkowski C., Fontanelli P. 1987, A\&AS 69, 263
\bibitem[1999]{chiaberge99}
Chiaberge, M., Capetti, A., Celotti, A., 1999, A\&A, 349, 77 
\bibitem[]{}Chiaberge M., Macchetto D.F., Sparks W.B., Capetti A., Allen M.G.,
Martel A.R., 2002, ApJ 571, 247


\bibitem[Conway \& Blanco(1995)]{con95}Conway J.E.,  Blanco P.R. 1995, ApJ, 449, 131
\bibitem[]{} Conway J., 1996 in {\it IAU 175, Extragalactic radio sources},  Edited by Ron D. Ekers, C. Fanti, and L. Padrielli. Published by Kluwer Academic Publishers, p. 92.
\bibitem[]{} Conway J., 1999 New Astronomy Review 43, 509
\bibitem[]{} Cotton W.D., Feretti L., Giovannini G., Lara L., Venturi T.
1999, Apj 519,108
\bibitem[]{} Croton D. et al. 2006 MNRAS, 365, 11
\bibitem[]{} de Young D., Roberts  M.S., Saslaw W.C. 1973, ApJ 185, 809
\bibitem[]{} Dressel, L. L.; Davis, M. M.; Bania, T. M. 1983,  ApJ 266, L97
\bibitem{} Emonts B.H.C., Morganti, R., Tadhunter, C. N., et al. 2006 A\&A 454, 125
\bibitem{} Emonts, B.H.C. 2006, PhD thesis, University of Groningen 
\bibitem{} Forbes D.A., 1991 MNRAS 249, 779
\bibitem[]{} Garcia A.M. 1993, A\&ASuppl.Ser. 100,47
\bibitem[]{}Gupta N., Saikia D.J. 2006, MNRAS 370, 738
\bibitem[]{} Heckman T.M., Balick B., van Breugel W.J.M., Miley G.K. 1983, AJ 
88, 583
\bibitem[]{} Heinz, S., Merloni, A., Schwab, J., 2007, ApJ  658, L9
\bibitem[Holt et al.(2006)]{ho06} Holt J., Tadhunter C., Morganti R., Bellamy M., Gonz‡lez Delgado R. M., Tzioumis A., Inskip K. J. 2006 MNRAS 370, 1633
\bibitem[]{} Holt J., Tadhunter C.N., Morganti R., 2008, MNRAS 387, 639 
\bibitem[]{} Laing R.A., Canvin J.R., Cotton W.D., Bridle A.H. 2006  MNRAS 368, 48 
\bibitem[]{} Leon S., Lim J., Combes F., Dihn-V-Trung 2003, in {\sl Active Galactic Nuclei: from Central Engine to Host Galaxy} Collin S., Combes F., Shlosman I. eds. ASP p.525
\bibitem[]{} Knapp G.R., Bies W.E., van Gorkom J.H. 1990, AJ 99, 476
\bibitem[]{} Mack K.-H., Klein U., O'Dea C.P., Willis A.G., Saripalli L. 1998, A\&A 329, 431

\bibitem[]{}Matsumoto Y., Fukazawa K., Iyomoto N., Makishima K. 2001,
Publ. Astr. Soc. Japan 53, 475
\bibitem[]{} Mirabel I.F. 1990, ApJ 352, L37

\bibitem[]{}Morganti et al. 2001, MNRAS 323, 331 
\bibitem[]{}  Morganti R. 2002, in {\sl Issues in Unification of
AGNs}, Maiolino R., Marconi A., Nagar N.  eds., ASP Conf.  Series 258, 63
(astro-ph/0109056)
\bibitem[]{} Morganti, R. et al. 2004, A\&A 424, 119
\bibitem[]{} Morganti, R., Tadhunter, C.N., Oosterloo, T.A., 2005, A\&A, 444, L9
\bibitem[]{} Morganti, R., Emonts B., Holt J., Tadhunter C., Oosterloo T., Struve C. 2009 AN 330, 233
\bibitem[]{} Noel-Storr J., Baum S.A., Verdoes Kleijn G. et al. 2003, ApJS 148, 419
\bibitem[]{}Peck, A.  B.  1999, {\it PhD Thesis}, New Mexico Inst.  of Mining
and Technology;
\bibitem[]{}Peck, A. B., Taylor, G. B. \& Conway, J. E., 1999, ApJ, 521, 103
\bibitem[]{}Peck, A. B. \& Taylor, G. B., 2001, ApJ, 554, L14
\bibitem[]{}de Ruiter et al. 2002 A\&A  396, 857
\bibitem[]{}Salom\'e P. et al. 2006, A\&A 454, 437
\bibitem[Sault, Teuben \& Wright(1995)]{sault95}Sault, R.J., Teuben, P.J., Wright, M.C.H. 1995, in {\sl ``Astronomical Data Analysis Software and Systems IV''}, eds. R. Shaw, H.E. Payne and J.J.E. Haynes, ASP Conf. Series, 77, 433
\bibitem[]{} Shostak G.S., van Gorkom J.H., Ekers R.D., Sanders R.H., Goss
W.M., Cornwell T.J. 1983, A\&A 119, L3
\bibitem[]{}Taylor, G. B., O'Dea, C. P., Peck, A. B. \& Koekemoer, A. M., 1999, ApJ, 512, L27
\bibitem[]{} Trager S.C., Faber S.M., Worthey G., Jes\'us Gonz\'alez J. 2000, AJ 119, 1645

Schilizzi R.T. 2000, A\&A 354, 45
\bibitem[]{}van Gorkom, J. H., Knapp, G. R., Ekers, R. D., Ekers,
D. D., Laing, R. A., Polk, K. S. 1989, AJ 97, 708;
\bibitem[Vermeulen et al.(2003)]{ver03}Vermeulen, R. et al. 2003, A\&A, 404, 861
\bibitem[]{}Wakker B., Oosterloo T.A., Putman M. 2002 AJ 123, 1953
\bibitem[]{} Wilson A.  1996, in {\sl `` Energy Transport in radio galaxies and
quasars''}, Hardee, Bridle, Zensus eds., ASP Conf.  Series p.9
\bibitem[]{} Worrall D. M., Birkinshaw M., Hardcastle M. J., 2001, MNRAS, 326, L7 
\bibitem[]{} Worrall D. M., Birkinshaw M., Hardcastle M. J., 2003, MNRAS, 343, L73   
\end{thebibliography}
\end{document}